\documentstyle{article}
\title{Renormalization in quantum Brans-Dicke gravity}
\author{ Z. Haba\\Institute of Theoretical Physics, University of Wroclaw,
\\50-204 Wroclaw, Plac Maxa Borna 9,Poland\\e-mail:zhab@ift.uni.wroc.pl}
\date{}
\begin{document}
\maketitle
\begin{abstract}
In the Brans-Dicke model we treat the scalar field exactly and
expand the gravitational field in a power series. A comparison
with  2D sigma models and $\phi^{4}$ perturbation theory in four
dimensions suggests that the perturbation series
 in 4D Brans-Dicke model is renormalizable.

\end{abstract}

\section{Introduction}
The appearance of a dimensional coupling
constant in Einstein gravity supplies  a
 simple argument of non-renormalizability of Einstein
 quantum gravity. The Brans-Dicke model \cite{brans}
 with the action
\begin{equation}
 W=\int dx(\frac{1}{2\xi}\sqrt{g}g^{\mu\nu}\partial_{\mu}\phi\partial_{\nu}\phi
 -\frac{1}{2}\sqrt{g}\phi^{2} R )
\end{equation}
has been constructed in such a way that a scalar
 field appears in the place of the coupling constant.
 In such a case  there is no dimensional coupling constant
 in the  model. Unfortunately, in the Brans-Dicke
 Lagrangian (1) there is no free Lagrangian
 (quadratic in the metric tensor)  which
 could be a starting point for a perturbative expansion.
 Our basic idea is to do  the functional
  integration with  an action which  depends on the
  quantum scalar  field but is quadratic in the metric field.

 Another approach is discussed in \cite{deser}-\cite{hooft}.
  By a conformal rescaling of the   metric tensor in the
classical action (1) $g_{\mu \nu}\rightarrow \phi^{-2}g_{\mu\nu}$
we have $\sqrt{g}\phi^{2}R\rightarrow \sqrt{g}R$. Then, the
authors can perform the perturbation expansion around a free
action which is quadratic in the scalar field and (separately) in
the metric field. It comes out to be non-renormalizable. However,
the latter approach is not equivalent to the first one. The
conformal (Weyl) invariance is lost at the quantum level because
of the quantum anomalies. As a consequence  Brans and Dicke's
main idea of dimensionless coupling is lost in such an approach.
Let us note that the loop expansion discussed in
\cite{deser}-\cite{hooft} when applied to the Brans-Dicke model
cannot be considered as the saddle point method  (where $\hbar$
is to be the small parameter). In $W/\hbar$ in eq.(1) the
parameter $\hbar$ can be absorbed into $\phi$ and there remains
no small  parameter which could  serve for an expansion in both
the metric and the scalar fields (we show  however that our
perturbation expansion in the metric field can be considered as
an expansion in $\xi^{-\delta}$ with a certain $\delta>0$). Let
us note that there is some similarity between our  functional
integration over $\phi$ and a preliminary integration over
longitundinal modes of the metric advocated in \cite{hawking}
-\cite{schleich}.

The Brans-Dicke model appears as the 1-loop  approximation to the
string theory \cite{tseytlin}\cite{cal}\cite{vene}. In such a
case the two approaches mentioned above are called the "string
frame" and the "Einstein frame". From our point of view such a
distinction is essential. This is the "string frame" applied at
short distances which requires a renormalization. The "Einstein
frame" whose quantum version is discussed in
\cite{deser}-\cite{hooft} applies to long distances and therefore
can remain unrenormalized.
 It follows from our results that there is no
need  for the string tension parameter to serve  as an
ultraviolet cutoff for  the effective string  field theory.
However, this dimensional parameter can distinguish the standard
free behaviour of renormalizable quantum field theories from a
regular behaviour on  sub-planckian distances \cite{habalet}. In
a  derivation of the effective string field theory only  the
massless modes are considered \cite{massless}. For this reason
the model is scale invariant. The exact theory should contain
mass parameters breaking the scale  invariance.

The classical  Brans-Dicke model of scalar-tensor gravity is scale
invariant,
i.e.,
$ W(\tilde{g},\tilde{\phi})=W(g,\phi)$ with
  \begin{equation}
  \tilde{g}_{\mu\nu}(x)=g_{\mu\nu}(\lambda x)
  \end{equation}
  and
  \begin{displaymath}
  \tilde{\phi}(x)=\lambda\phi(\lambda x)
  \end{displaymath}
In order to preserve positivity of the metric $g$ we write
\begin{equation}
\sqrt{g}g^{\mu\nu}=(\exp h)^{\mu\nu}\equiv G^{\mu\nu}
\end{equation}
where $h^{\mu\nu}$ is also dimensionless (in the classical model)
and assumed to be small. Eq.(3) seems to be a reasonable starting
point for an expansion around the flat space.
 However, in the conventional
perturbation expansion of Einstein quantum gravity we would
encounter a difficulty with the assumptions (2)-(3) because
already at the lowest order $h$ has a scale dimension 1. We show
that this obstacle can be avoided in the Brans-Dicke model. We
obtain a perturbation expansion which is consistent with the
scaling (2).

We shall give arguments showing that the ultraviolet behaviour of
the $4D$ Brans-Dicke  theory is similar to the ultraviolet
behaviour of (non-compact) $\sigma$- models in two dimensions.
For this aim let us write down the action (1) in a way resembling
an action of $GL(4,R)$-valued sigma field
\cite{capper}-\cite{strominger}.
 So, we treat $G$ in eq.(3) as an
element of $ GL(4,R)$. We can represent the Christoffel symbol in
the form (see ref.\cite{capper}, Appendix A)
\begin{equation}
\Gamma=a J+b G^{-1} J G
\end{equation}
where
\begin{displaymath}
J_{\mu}=G^{-1}\partial_{\mu} G
\end{displaymath}
Here, $J$ is $4\times 4$ matrix; in eq.(4) the components
of $\Gamma$ are expressed by matrix elements of $J$ and
$G$ in a way which is irrelevant at the moment.
 Now, in $ W(G)=\frac{1}{2} \int dx
\phi^{2}G^{\mu\nu}R_{\mu\nu} $ the Ricci tensor is of the form
$\partial \Gamma-\Gamma\Gamma$. Hence, the final formula for the
gravitational action $W(G)$ has the
form resembling the $\sigma$-model
\begin{equation}
\begin{array}{l}
W(G)=\int dx\phi^{2}G(a_{1}JJ +a_{2}\partial J)
\end{array}
\end{equation}
where $a_{k}$ are certain constants. The precise contraction of
indices is rather complicated and irrelevant for our discussion.
The matrix notation is useful when we  wish to follow the
perturbation expansion. So, the lowest order in the expansion of
$W(G)$ in terms of $h$ is  $W_{1}(G)= \int\phi^{2}\partial^{2}h $.
 Then, there is the bilinear term $\frac{1}{2}\phi^{2}h\partial^{2}
h$ which will be considered as the free action $W_{2}(G)$. Now,
an expansion of the interaction $W_{I}\equiv
W(G)-W_{1}(G)-W_{2}(G)$ starts as follows
\begin{equation}
W_{I}(G)= \int dx\phi^{2}(b_{1} h\partial h\partial h
+b_{2}hh\partial^{2}h+b_{3}hh\partial h\partial h+
b_{4}hhh\partial^{2}h+.....)
\end{equation}
with certain constants $b_{k}$.

\section{The lowest order }
We work in a generalization of the harmonic gauge
\begin{equation}
\partial_{\mu}(\phi^{2}G^{\mu\nu})=0
\end{equation}
This gauge has the advantage that the bilinear form quadratic in
the second order derivatives has an exceptionally simple form.
So, the gravitational part $W(G)$ of the action (1) in the gauge
(7) can be written as
\begin{equation}
W(G)=\frac{1}{4}\int dx
(\phi^{2}_{,\nu}G^{\mu\nu}\Gamma^{\sigma}_{\mu\sigma}
-\phi^{2}_{,\sigma}G^{\mu\nu}\Gamma^{\sigma}_{\mu\nu}
+\phi^{2}G^{\mu\nu}\Gamma^{\sigma}_{\mu\nu,\sigma})
\end{equation}
Then, the quadratic form which is of the second order in
 derivatives reads
\begin{equation}
W_{2}^{Q}(G) =-\frac{1}{8}\int dx \phi^{2}\hat{h}^{\alpha\beta}\triangle
\hat{h}^{\alpha\beta} +\frac{1}{32}\int dx\phi^{2}H\triangle H
\end{equation}
where  $\triangle$ is the Laplace operator in $R^{4}$,
\begin{displaymath}
\hat{h}^{\alpha\beta}=h^{\alpha\beta}-\delta^{\alpha\beta}\frac{1}{4}H
\end{displaymath}
is the traceless part of $h$ and $H=Tr h=h^{\alpha\alpha}$ denotes the
trace of $h$.
 Then,
 the bilinear form resulting from $W$ together with the Faddeev-Popov determinant
 for the gauge (7) define the
 integration measure
\begin{displaymath}
d\mu(h,\phi)={\cal D} h{\cal D}
\phi\exp\left(-W_{2}\left(h,\phi\right)\right)\det D_{FP}
\end{displaymath}
where
\begin{equation}
W_{2}(h,\phi)=\frac{1}{2\xi}\int dx\phi (-\triangle)\phi +
\frac{1}{4}\int dx
\hat{h}^{\mu\nu}(-\triangle_{2T})\hat{h}^{\mu\nu}
-\frac{1}{16}\int dxH(-\triangle_{2S})H
\end{equation}
The Faddeev-Popov  operator for the gauge (7) is
\begin{equation}
D_{FP}=\frac{1}{2}\phi^{2}G^{\mu\nu}\partial_{\mu}\partial_{\nu}
\end{equation}
The operators in the quadratic form in $W_{2}(G)$ can be deduced
from eqs.(8)-(11). So,
\begin{equation}
\triangle_{2T}=\frac{1}{2}\phi^{2}\triangle+\frac{1}{2}\phi^{2}_{,\nu}C^{\nu\mu}\partial_{\mu}
\end{equation}
Here, the operator $\triangle_{2T} $ is acting on tensors of the
second order, the Laplacian $\triangle$ is considered as the
diagonal operator in the space of tensors and $C^{\mu\nu}$ are
matrices built from $\delta$-functions whose explicit form can be
read from eq.(8). The operator $\triangle_{2S}$ is acting on
scalars and has the form
\begin{equation}
\triangle_{2S}=\frac{1}{2}\phi^{2}\triangle-\frac{1}{4}\phi^{2}_{,\nu}\partial_{\nu}
\end{equation}
 We can generalize the model (10) for an application of the
dimensional regularization replacing the scalar free action
$W_{2}(\phi)=\frac{1}{2\xi}\int dx \phi(-\triangle)\phi$
 by
\begin{equation} W_{2}^{\gamma} (\phi)=\frac{1}{2\xi}\int
dx\phi (-\triangle)^{2-\gamma}\phi
\end{equation}
The resulting Brans-Dicke action will be denoted $W^{\gamma}$ .
The action $W^{\gamma}$ is also scale invariant
$W^{\gamma}(\tilde{h},\tilde{\phi})= W^{\gamma}(h,\phi) $ where
$\tilde{\phi}(x)=\lambda^{\gamma}\phi(\lambda x)$ and
$\tilde{h}(x)=\lambda ^{1-\gamma}h(\lambda x)$.

For a perturbation expansion instead of the determinant $\det
D_{FP}$ the Faddeev-Popov ghosts $(\overline{\chi},\chi)$ are
introduced. The corresponding action is
\begin{equation}
W(\chi)=\frac{1}{2}\int dx
\phi^{2}G^{\mu\nu}\overline{\chi}\partial_{\mu}
\partial_{\nu}\chi
\end{equation}
It is scale invariant, i.e., invariant under the change of
variables $\tilde{\chi}(x)= \lambda ^{1-\gamma}\chi(\lambda x)$
(together with the scale change of $\phi$ and $h$ mentioned
before).

We apply  some  methods of our earlier paper \cite{habalet} where
an interaction with a scale invariant metric has been discussed.
First, let us consider the propagator for the Faddeev-Popov ghosts
(15) in the proper time representation.
We take an average over $\phi$ of the heat kernel $K_{\tau}$
of the operator $D_{FP}$. The path integral formula takes
the form
\begin{equation}
\begin{array}{l}
\langle
K_{\tau}(x,y)\rangle=\langle(\exp\tau D_{FP})(x,y)\rangle=
\int {\cal D}\phi\exp(-W_{2}(\phi) )\cr \int {\cal
D}q\exp(-\frac{1}{2}\int \phi^{2}G^{\mu\nu}\frac{dq_{\mu}}{dt}
  \frac{dq_{\nu}}{dt})
 \delta\left(q\left(0\right)-x\right)
  \delta\left(q\left(\tau\right)-y\right)
  \end{array}
  \end{equation}
 In order to obtain the behaviour of $\langle K_{\tau}(x,y)\rangle $
 for a small $\tau$ it is  sufficient to assume the scale invariance
 of  $\phi$
 \begin{equation}
 \int dx {\cal L}\left(\lambda^{\gamma}\phi\left(\lambda
 x\right)\right) =\int dx {\cal L}\left(\phi\left(x\right)\right)=
 W_{2}(\phi)
 \end{equation}
   In fact,  introducing in eq.(16) a new functional integration
   variable $\tilde{\phi}=  \lambda^{\gamma}\phi\left(\lambda
 x\right)              $ and a new path  $\tilde{q}$
 defined on an interval $[0,1]$ by
 \begin{equation}
q(\tau s)-x= \tau^{\sigma}\sqrt{\xi}(\tilde{q}(s)-x)
\end{equation}
where
 \begin{displaymath}
 \sigma=\frac{1}{2}(1+\gamma)^{-1}
 \end{displaymath}
 we can explicitly extract the $\tau$-dependence of
 $\langle K_{\tau}(x,y)\rangle $
 \begin{equation}
 \begin{array}{l}
 \langle K_{\tau}(x,y)\rangle=\langle E[\delta\left(y-x-
 \tau^{\frac{1}{2}-\sigma\gamma}\sqrt{\xi}\tilde{q}\left(1\right)\right)]\rangle
 \cr
 =\xi^{-2}\tau^{-\frac{2}{1+\gamma}}F\left(\tau^{-\frac{1}{2(1+\gamma)}}
 \xi^{-\frac{1}{2}}\left(y-x\right)\right)
 \end{array}
 \end{equation}
 where $E[.]$ denotes an expectation value over the paths
 $\tilde{q}$ and $F$ is the probability
 distribution of $\tilde{q}(1)$  .
  Then, an average of the propagator
 (in the proper time representation) behaves as
 \begin{equation}
 \begin{array}{l}
  \langle (-D_{FP})^{-1}(x,y) \rangle =\int_{0}^{\infty}
  d\tau \langle K_{\tau}(x,y)\rangle
  \cr
  =\xi^{-1-\gamma}\int_{0}^{\infty}d\tau \tau^{-\frac{2}{1+\gamma}}F\left(\tau^{-\frac{1}{2(1+\gamma)}}\left(y-x\right)\right)
  \simeq \xi^{-1-\gamma}\vert x-y\vert^{-2 +2\gamma}
  \end{array}
  \end{equation}
if $\gamma<1$. For $\gamma=1$
\begin{equation}
\langle (-D_{FP})^{-1}(x,y)\rangle \simeq -\xi^{-2}\ln(x-y)^{2}
\end{equation}
We can represent the Faddeev-Popov determinant by its heat kernel
\begin{equation}
\det D_{FP}=\exp Tr\ln D_{FP}=
\exp\left(-\int_{0}^{\infty}\frac{d\tau}{\tau} \int dx
\left(\exp\left(\tau D_{FP}\right)-1\right)\left(x,x\right)\right)
\end{equation}
It follows from the small $\tau$ behaviour (19) of the heat kernel
(16) that the integral over the proper time in eq.(22) is finite.
Hence, no renormalization
 is needed for the Faddeev-Popov determinant .

We consider next the heat kernel $K_{\tau}$  and the propagator
of the operator $\triangle_{2T}$. The path integral formula takes
the form
\begin{equation}
\begin{array}{l}
\langle (-\triangle_{2T})^{-1}(x,y)\rangle=\int_{0}^{\infty}d\tau
\langle K_{\tau}(x,y)\rangle=\int_{0}^{\infty}d\tau
\langle(\exp\tau\triangle_{2T})(x,y)\rangle\cr
=\int_{0}^{\infty}d\tau\int {\cal D}\phi\exp(-W_{2}(\phi) ) \int
{\cal D}q\exp(-\frac{1}{2}\int \phi^{2}\frac{dq_{\mu}}{dt}
  \frac{dq^{\mu}}{dt})T_{\tau}
 \delta\left(q\left(0\right)-x\right)
  \delta\left(q\left(\tau\right)-y\right)
  \end{array}
  \end{equation}
 where the matrix $T_{\tau}$ is the solution of the equation
  \begin{equation}
  dT_{\tau}=\phi_{,\mu}C^{\mu\nu}T_{\tau} dq_{\nu}
  \end{equation}
  and the  stochastic
  integral in eq.(24) is of the Ito type
  (see ref. \cite{ikeda} ).
  The formula (23) is analogous to the well-known Feynman formula
  for  a particle in an electromagnetic field. We obtain a similar
  formula for the heat kernel of $\triangle_{2S}$.
  The argument with the change of
   variables (17)-(18) applies well
   to the functional integral (23). The change of time in
   $T_{\tau}$ is irrelevant for the short time behaviour
   of the $\tau$-integral in  (23) because $T_{\tau}$ is continuous in $\tau$.
  Hence, the argument used at the propagator for ghosts
  applies here as well leading to the result
 \begin{equation}
 \begin{array}{l}
  \langle (- \triangle_{2T})^{-1}(x,y) \rangle =\int_{0}^{\infty}
  d\tau \langle K_{\tau}(x,y)\rangle
  \cr
  \simeq \xi^{-1-\gamma}\vert x-y\vert^{-2+2\gamma}
  \end{array}
  \end{equation}
if $\gamma<1$ and for $\gamma=1$  we obtain the behaviour (21). It
is clear that also
\begin{displaymath}
\langle (- \triangle_{2S})^{-1}(x,y) \rangle
  \simeq \xi^{-1-\gamma}\vert x-y\vert^{-2+2\gamma}
  \end{displaymath}
  Strictly speaking the Euclidean functional integral over $H$
  is divergent. It has to be interpreted by means of a contour
  rotation  in the functional space of metrics, see
  \cite{hawking}\cite{mazur}.
\section{A perturbation expansion}
The functional integral  can  be expressed in the form
\begin{equation}
d\mu_{0}(\phi)d\nu_{T}(\hat{h})d\nu_{S}(H)\det(-\triangle_{2T})^{-\frac{1}{2}}
\det(-\triangle_{2S})^{-\frac{1}{2}}\det D_{FP}
\end{equation}
where $\mu_{0}$ is a Gaussian measure with the covariance
$\xi(-\triangle)^{-1}(x,y)$ , $\nu_{T}$ is a  Gaussian measure
with the covariance $(-\triangle_{2T})^{-1}(x,y)$ and $\nu_{S}$
is a Gaussian measure with the covariance
$(-\triangle_{2S})^{-1}(x,y)$. The determinants of
$\triangle_{2T}$ and $\triangle_{2S}$ require no renormalization
by the same argument as the one used in the discussion of $\det
D_{FP}$.
 Eq.(21) means that
\begin{equation}
\int d\mu_{0}(\phi)\int d\nu_{T}(\hat{h})
\hat{h}_{\mu\nu,\rho}(x)\hat{h}_{\alpha\beta,\sigma}(y)\simeq
-\xi^{-2}\partial_{\rho}
\partial_{\sigma}
\ln(x-y)^{2}
\end{equation}
The higher order correlations are expectation values over $\phi$
of sums of products of $(-\triangle_{2})^{-1}$
\begin{equation}
\begin{array}{l}
\langle \hat{h}(x_{1})......\hat{h}(x_{2n})\rangle=
\sum_{(i,j)}\langle
\prod_{(i,j)}(-\triangle_{2T})^{-1}(x_{i},x_{j})\rangle
=\int_{0}^{\infty}d\tau_{1}....d\tau_{n}\cr \langle
E[T_{\tau_{1}}\delta\left(x_{2}-x_{1}-
 \tau_{1}^{\frac{1}{2}-\sigma\gamma}\sqrt{\xi}\tilde{q}_{1}\left(1\right)\right)
 ........T_{\tau_{n}}\delta\left(x_{2n}-x_{2n-1}-
 \tau_{n}^{\frac{1}{2}-\sigma\gamma}\sqrt{\xi}\tilde{q}_{n}\left(1\right)\right)
 ]\rangle
\cr +permut.
\end{array}
\end{equation}
where on the r.h.s. we have a sum over permutations of $x_{k}$ in
accordance with the well-known Gaussian combinatorics. It can be
shown that the singularity of an expectation value of a product
of $(-\triangle_{2T})^{-1}$ is the same as that of a product of
singularities of $\langle (-\triangle_{2T})^{-1} \rangle$. It is
clear that when we are applying  formally the propagator (16) for
the $h$-average then an expansion in $h$ has the same singularity
as an expansion in the two-dimensional $\sigma$- model. The
latter is known to be renormalizable. In the Brans-Dicke model
the problem is more complicated because in the perturbation
expansion we have the mixed $\phi h$ terms which lead to
additional singularities. However, it can be seen from
eqs.(5)-(6) that the first term in eq.(5) contributes to the
interaction $W_{I}$ terms of the form $\psi^{2}\phi^{2}$ times
powers of a field $h$ of zero dimension where $\psi=\partial h$
and $\phi$ are fields of dimension 1. Hence, by power counting
the perturbation expansion has the same singularity as the one in
the $\phi^{4}$  model (or a $\phi^{2}\psi^{2}$ model). The second
term in eq.(5) gives in $W_{I}$ terms of the form $\phi^{2}B$
times fields of dimension zero where $B=\partial\partial h$ is a
field of dimension 2. The singularity of such  an interaction is
of the same kind as the one coming from the $\phi^{2}\psi^{2}$
interaction. From the part of the Lagrangian (1) with derivatives
of the scalar field we obtain an interaction of the form
$W_{I}(\phi)=\int dx (G-1)\partial \phi\partial\phi$. A
renormalization of the perturbation expansion of this interaction
is equivalent to the well-studied problem of the renormalization
of a composite field (the energy-momentum tensor) in the  free
massless scalar quantum  field theory. We have shown in sec.2
that the Faddeev-Popov determinant requires no renormalization.
However, if we expand the action (15) around $G=1$ then we obtain
the same problem of a renormalization of the composite field as
the one concerning the scalar field.

Calculating the terms in the perturbation expansion explicitly we
can see that  there are quartically divergent vacuum diagrams, as
for example (with certain powers $n$ and $m$)
\begin{displaymath}\int dxdy\langle \phi^{2}(x)h^{n}(x)\partial h(x)\partial
h(x)\phi^{2}(y) h^{m}(y)\partial h(y)\partial h(y)\rangle
\end{displaymath}
which are analogs of $\int dxdy \langle
\phi^{2}(x)\psi^{2}(x)\phi^{2}(y)\psi^{2}(y)\rangle$
in the $\phi^{2}\psi^{2}$ model.

We can check renormalizability of connected diagrams (and the
correspondence to the $\phi^{2}\psi^{2}$ model) at lower orders
again by explicit calculations. So, let us consider the two-point
function  $\langle h(x_{1})h(x_{2})\rangle$ in the second order
of $W_{I}$
\begin{equation}
\begin{array}{l}
\int dxdy\langle :\phi^{2}(x)h^{n}(x)\partial h (x)\partial
h(x): :\phi^{2}(y)h^{m}(y)\partial h (y)\partial h(y):
h(x_{1})h(x_{2}) \rangle
\end{array}
\end{equation}
From $\langle \phi^{2}\phi^{2}\rangle $ we obtain $(y-x)^{-4}$
and from $\langle\partial h(x)\partial h(y)\rangle$ we have
$(x-y)^{-2}$. So, the connected diagrams are quadratically divergent.
This is an analog of the diagram
\begin{displaymath}
\langle:\phi^{2}(x)\psi^{2}(x)::\phi^{2}(y)\psi^{2}(y):\psi(x_{1})
\psi(x_{2})\rangle
\end{displaymath}
which leads to the $\psi$-mass renormalization
    in the $\phi^{2}\psi^{2}$  model ( this will be  the
   wave function renormalization in the model (5)).
The four-point function at the same order would be logarithmically
divergent. The divergence could be cancelled by the coupling
constant renormalization (of the $\phi^{2}\psi^{2}$ interaction).
Higher order connected correlation functions are finite at the
second order of $W_{I}$.

                            We could continue with higher orders.
 So, the third order term in  the interaction leads
 to  the higher order wave function renormalization in the propagator
\begin{equation}
\begin{array}{l}
\int dy_{1}dy_{2}dy_{3}dy_{4} \langle
\phi^{2}(y_{1})h^{n}(y_{1})\partial h (y_{1})\partial h(y_{1})
\phi^{2}(y_{2})h^{m}(y_{2})\partial h (y_{2})\partial h(y_{2}) \cr
\phi^{2}(y_{3})h^{r}(y_{3})\partial h (y_{3})\partial h(y_{3})
   h(x_{1})h(x_{2})\rangle
   \end{array}
\end{equation}
In general, we can see that the expansion in $h$ can be
considered as an expansion in powers of
$\xi^{-\frac{1}{2}-\frac{\gamma}{2}}$.

A complete proof of the renormalizability still requires  a more
detailed analysis but we think that we have given here some
convincing arguments showing that a functional integration over
the scalar field substantially improves the ultraviolet behaviour
of the perturbation series. Another interesting question concerns
the scale invariance of the summed up
 renormalized series. Although at each order the metric
 field $h$ has zero dimension the logarithms can sum up
 in $G=\exp h$ leading to a positive  dimensionality
 of the metric field $G$.
In fact,  if we treat the matrix $h$ like a commutative Gaussian
field then
\begin{equation}
\int d\mu_{0}(\phi)\int d\nu_{T}(\hat{h}):\exp \hat{h}(x): :\exp
\hat{h}(y):\simeq \exp \frac{1}{2} \langle
(-\triangle_{2T})^{-1}(x,y)\rangle
\end{equation}
The scale dimension of the metric (if positive) determines the
short distance behaviour of matter fields \cite{habalet}. If the
approximation (31) is reliable then a determination of the
constants in eq.(21) would determine the short diststance
behaviour of all quantum fields interacting with gravity.

\end{document}